# OPTIMIZING THE SEARCHES FOR INTERSTELLAR HETEROCYCLES

Emmanuel E. Etim*[1], ROA Adelagun[1], Chrysanthus Andrew[1] and Oladimeji Enock Oluwole[2,3]

*email:emmaetim@gmail.com

[1]Department of Chemical Sciences, Federal University Wukari, Taraba State, Nigeria

[2]Department of Physical Sciences , Joseph Ayo Babalola University, Ikeji-Arakeji, Nigeria.

[3]The Institute for Physical Research and Technologies, Peoples' Friendship University of Russia, Moscow, Russia.

**Abstract**: It is a fact that interstellar formation processes are thermodynamically affected. Based on this, the seven heterocycles; imidazole, pyridine, pyrimidine, pyrrole, quinoline, isoquinoline and furan that have been searched for from different astronomical sources with only upper limits of their column density determined without any successful detection remain the best candidates for astronomical observation with respect to their isomers. These molecules are believed to be formed on the surface of the interstellar dust grains and as such, they are susceptible to interstellar hydrogen bonding. In this study, a two way approach using ab initio quantum chemical simulations is considered in optimizing the searches for these molecules in interstellar medium. Firstly, these molecules and their isomers are subjected to the effect of interstellar hydrogen bonding. Secondly, the deuterated analogues of these heterocycles are examined for their possible detectability. From the results, all the heterocycles except furan are found to be strongly bonded to the surfaces of the interstellar dust grains thereby reducing their abundances, thus contributing to their unsuccessful detection. Successful detection of furan remains highly feasible. With respect to their D-analogues, the computed Boltzmann factor indicates that they are formed under the dense molecular cloud conditions where major deuterium fractionation dominates implying very high D/H ratio above the cosmic D/H ratio which suggests the detectability of these deuterated species.

## 1. INTRODUCTION

The interstellar medium (ISM) remains a unique laboratory. Most of the chemical species that exist in it are either too difficult to be found or do not exist at all in the terrestrial laboratory. These molecules thus serve as guides for new chemistry in the terrestrial laboratory. To the astronomers and astrophysicists, these molecules are perfect probes of astrophysical phenomena while to the astrochemists, they provide information to better understand the chemistry and chemical composition of the ISM. Using these molecules to shed more light on the chemical origin of life and how the small chemical species that were present on the early may have given birth to the complex systems and structures that we see today are a major concern to the astrochemists. The importance of these molecules coupled with the curiosity of understanding the solar system and in general the world around have resulted in the unique detection and characterization of over 200 of these molecules largely via their rotational transition spectra.[1-3] Among the known interstellar and circumstellar molecules, isotopologues of H, C, O, N and S-containing molecules have also been detected. Just as H-containing molecules constitute majority of all the known astrophysical molecules, deuterated molecules are the highest known interstellar and circumstellar isotopologues.[4]

Heterocycles are major players in different aspects of terrestrial biology. The nitrogen containing heterocycles are fundamental building blocks for nucleic acid. Their applications in medicine, industry and other areas are well recognized.[5-8] Nitrogen heterocycles, amino acids, carboxylic acids, sugar related compounds, phosphorus compounds, sulphur compounds, amine and amides, aliphatic and aromatic hydrocarbon, and macromolecular materials like polyaromatic hydrocarbons (PAHs) are among the various classes of molecules that have been detected in carbonaceous meteorites.[7] The detection of these molecules from carbonaceous meteorites demonstrates their possible formation and survival outside the Earth.[5,7,9-10] The possible origin of the molecules and/or their precursors has been linked to the interstellar or circumstellar medium. Branched chain molecules are very common among the materials detected in carbonaceous materials; the recent detection of a branched chain molecule in ISM (iso-propyl cyanide) further stresses the link between the interstellar medium and the molecular composition of meteorites.[11-13] All of these have prompted the astronomical searches for heterocycles.

Seven heterocycles; imidazole, pyridine, pyrimidine, pyrrole, quinoline, isoquinoline and furan that have been searched for from different astronomical sources with upper limits in the range of $4*10^{12}$ to $2.8*10^{21}$cm$^{-2}$ determined for their column densities in all the cases without any successful detection.[14-21] Interstellar formation processes are largely thermodynamically controlled. From the energy, stability and abundance (ESA) relationship existing among interstellar and circumstellar molecules, it has been shown that there is a direct link between the stability of a molecule and its interstellar abundance which influences its astronomical observation. For related molecules like isomers which are very common among the known interstellar and circumstellar molecules, it is observed that the higher the stability of an isomer, the higher its interstellar abundance which enhances its astronomical detection as compared to others with low stability. Concerned by the unsuccessful detection of all the heterocycles that have been astronomically searched, these heterocycles alongside their various isomers were subjected to stability check in line with the ESA relationship.

Interestingly, it turns out that these heterocycles are the best candidates for astronomical searches as they are found to be the most stable isomers in their respective isomeric groups.[3,4,22]

Processes that occur on the surface of the interstellar dust grains are the dominant processes for the formation of interstellar molecules over the gas phase processes. Due to its chemical composition, the surface of the interstellar dust grains serves as a platform for hydrogen bonding among the molecules that are formed on it. This interstellar hydrogen bonding leads to a reduction in the interstellar abundance of the molecule involved as a greater part of the molecules is attached to the surface of the interstellar dust grains. For instance, acetic acid is more stable than its isomer; methyl formate but in all astronomical sources, methyl formate is found to be highly abundant than acetic acid and the reason for this is traced to interstellar hydrogen bonding in which acetic acid is strongly bonded to the surface of the interstellar dust grains thereby reducing its abundance as compared to methyl formate.[31] This explains the exception in the ESA relationship where the most stable isomer is not the most abundant.[23]

Optimizing the searches for interstellar heterocycles is the aim of the present study. In doing this, a two way approach using ab initio quantum chemical calculation is considered. Firstly, these heterocycles and their isomers are subjected to the effect of interstellar hydrogen bonding which could affect the interstellar abundance of these heterocycles thus making their astronomical detection difficult. Secondary, a recent study has shown that subjecting large molecules of astrophysical interest like PAHs (which are similar to the heterocycles considered here) to interstellar medium conditions leads to the transformation of these molecules through hydrogenation, oxygenation and hydroxylation to complex organics. As a consequence, these fully hydrogenated molecules lose their spectroscopic signature which in turn contributes to their lack of astronomical detection.[24] This suggests the searches for the isotopologues of these molecules as a means of mitigating this effect. Based on this, the deuterated isotopologues of these heterocycles are examined. One of the main distinctions between a molecule and its deuterated analogue is the difference in the zero point energy (ZPE). From this difference, the Boltzmann factor which gives insight about the possible formation conditions for the deuterated molecule can be computed. This has been computed for all the heterocycles considered here in order to investigate the nature of deuterium fractionation and the possible detectability of these D-analogues

## 2. METHODOLOGY

Mimicking the interstellar laboratory conditions in our terrestrial laboratory could be an extremely difficult task especially when the system of interest involves short-lived species, intermediates and transitions states. However, such conditions and systems can be simulated using computational and theoretical tools. For the present study, Gaussian 09 suit of program is used for all computational simulations.[25] In order to obtain high accurate results that could be in good agreement with experimental data (if available) the Møller–Plesset second order perturbation theory; MP2 (full) with the 6-311++G** basis set was used in examining the interstellar hydrogen bonding while the G4 compound model was used in obtaining the ZPE for the heterocycles and their corresponding D-analogues. The choice of these methods came from previous experience.[26-31] The optimized structures used for all the results reported here were found to be stable with no imaginary frequency as confirmed by the harmonic vibrational frequency calculations. The binding energy (B. E) of the complex formed between the heterocycles and the water molecule on the surface of the interstellar dust grains is determined using the super molecular approach which is defined as:

$$B.E(complex) = E(complex) - [E(water molecule) + E(heterocycle molecule)] \quad (1)$$

The reported binding energies have been corrected for the basis set superposition error (BSSE) using the counterpoise method developed by Boys and Bernardi[33].

The difference in the ZPE between a molecule and its deuterated analogue expressed in terms of temperature (in Kelvin) represents the temperature at which the Boltzmann factor (E/kT) for the system is unity. Considering the temperature range of ≈10 to 100k typical of the dense molecular clouds where these heterocycles are believed to be formed, two temperature regimes of 10 and 100k are used in computing the Boltzmann factor for the systems considered here. The dipole moment of a molecule is among the factors considered for an astronomical observation of a molecule. This is because intensities of rotational transitions of a molecule scale with the square of the dipole moment, thus, the higher the dipole moment, the higher the intensity of the lines. As a result of this, dipole moment of these heterocycles and their corresponding isomers at the MP (full)/6-311++G** level are reported in all the cases considered here.

## 3. RESULTS AND DISCUSSION

Since the order of stability of these heterocycles and their isomers is known from previous study, for the effect of interstellar hydrogen bonding, the heterocycles and their three next most stable isomers are considered except in the $C_4H_5N$ and $C_9H_7N$ isomeric groups where their next to 5 and 6 most stable isomers are examined respectively. The effect of the substitution of deuterium at different positions on the ZPE of the deuterium molecules is also observed. The results for both approaches considered here for optimizing the searches for interstellar heterocycles are presented and discussed for each heterocycle.

### 3.1 Imidazole and Its Isomers

The three next most stable isomers of the $C_3H_4N_2$ group after imidazole are 3-aminoacrylonitrile, pyrazole and 2-aminoacrylonitrile respectively. The binding energy (B.E) of these isomers with water on the surface of the interstellar dust grains, their enthalpy of formation from previous study and their dipole moment are shown in Table 1.

**Table 1:** $C_3H_4N_2$ Isomers, Binding Energy with water, $\Delta H^o_f$ and Dipole moment

| Molecule | B. E with $H_2O$ (kcal/mol) | $^a\Delta H^o_f$ (kcal/mol) | Dipole moment (Debye) |
|---|---|---|---|
| Imidazole | -5.687 | 26.641 | 3.997 |
| 3-aminoacrylonitrile | -4.609 | 35.805 | 5.507 |
| Pyrazole | -6.205 | 36.958 | 2.546 |
| 2-aminoacrylonitrile | -3.734 | 43.493 | 4.391 |

$^a$ [ref. 30]

From the table, pyrazole is found to be more strongly bonded to the surface of the interstellar dust grains followed by the molecule of interest; imidazole while both 3-aminoacrylonitrile and 2-aminoacrylonitrile are the least bonded isomers. This implies that a reasonable portion of imidazole could be attached to the surface of the interstellar dust grains thereby reducing its overall interstellar abundance and hence, hampering its astronomical observation. Isomers are believed to have common precursors for their formation route, thus a successful detection of an isomer is an indication of the interstellar present of the other isomers not yet detected. This concept of isomerism has been an important search light for interstellar and circumstellar molecules with about 40% of all the known interstellar and circumstellar molecules having known isomeric counterparts. Thus, should the aminoacrylonitriles with high dipole moments, which are less bonded to the surface of the interstellar dust grains, be searched and possibly observed from the astronomical sources, this obviously will confirm the present and detectability of imidazole in space.

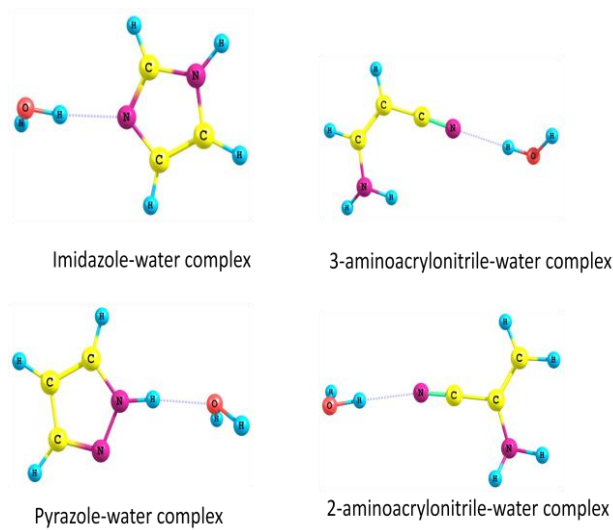

figure **1:** Optimized structures of the hydrogen bonded $C_3H_4N_2$ isomer complexes with water

Figure 1 pictures the optimized structures of the hydrogen bonded $C_3H_4N_2$ isomer complexes with water molecule considered here. Water molecule has the ability of acting as both hydrogen bond donor and acceptor. Both cases are observed here.

It acts as hydrogen bond acceptor in the case of pyrazole-water complex while in the other three complexes, it acts as hydrogen bond donor where the hydrogen atom in water is forming bond with the nitrogen atom of the respective $C_3H_4N_2$ isomer. There is an elongation of one of the O-H bonds from the original 0.959Å in the cases where the water molecule acts as hydrogen bond donor while both O-H bonds of water elongate in the case of pyrazole-water complex where it acts as hydrogen bond acceptor. This elongation of the O-H bond(s) of water shows evidence of hydrogen bond formation between the $C_3H_4N_2$ isomers and the water on the surface of the interstellar dust grains. Table 2 shows the ZPE of imidazole and its deuterated analogues, the Boltzmann factor (E/kT) at the two temperature regimes and the temperature (*ΔT) where the Boltzmann factor is unity.

Table 2: ZPE and Boltzmann factor (E/kT) for Imidazole and its D-analogues

| Molecule | ZPE (kcal/mol) | *ΔT (K) | E/kT (at | E/kT(at |
| --- | --- | --- | --- | --- |

|  |  |  | T=10k) | T=100k) |
|---|---|---|---|---|
| Imidazole | 44.645 | - | - | - |
| $C_3H_3D(6)N_2$ | 42.605 | 1026.602 | 102.660 | 10.266 |
| C3H3D(7)N$_2$ | 42.509 | 1075.232 | 107.523 | 10.752 |
| $C_3H_3D(8)N_2$ | 42.649 | 1004.813 | 100.481 | 10.048 |
| $C_3H_3D(9)N_2$ | 42.382 | 1139.020 | 113.902 | 11.390 |
| $C_3H_2D_2(6,7)N_2$ | 40.468 | 2102.150 | 210.215 | 21.021 |
| $C_3HD_3(6,7,8)N_2$ | 38.469 | 3108.226 | 310.823 | 31.082 |
| $C_3D_4N_2$ | 36.426 | 4136.091 | 413.609 | 41.361 |

*The difference in the ZPE of a molecule and its D-analogue expressed in terms of temperature.

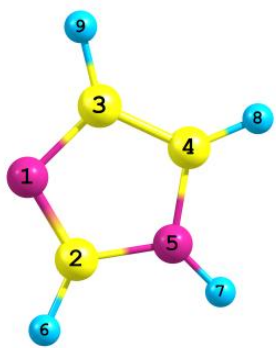

Figure 2: Optimized structure of imidazole

The number(s) in bracket in the first column of Table 2 represents the position of substitution of the deuterium atom on the imidazole molecule as shown in Figure 2. As would be expected, the ZPE of imidazole is much higher than that of its deuterated analogues. The four hydrogen atoms at positions 6, 7, 8 and 9 are all non-identical. Deuterium substitution at these positions also shows non-identical effects resulting in the different in the ZPE. With a Boltzmann factor of one, no major deuterium fractionation is expected to occur, thus the D/H ratio will reflect the cosmic or elemental D/H ratio of $\approx 10^{-5}$. From Table 2, it is obvious that having a Boltzmann factor of unity is not achievable within the conditions of the molecular clouds where these molecules are believed to form except they are formed in the hot cores with temperature range of 100 to 1000k. Under the conditions of the molecular clouds, the large value of the Boltzmann factor (E/kT at T=10 and 100k) indicates major deuterium fractionation which will result in high abundance of the deuterated analogue of imidazole since the D/H ratio will much higher than the cosmic D/H ratio. This of course signals the possible detectability of this D-analogue.

**3.2 Pyridine and Its Isomers**

As shown in Table 3, pyridine is by far the most stable isomer of the $C_5H_5N$ isomeric group. The other stable isomers after pyridine are also listed in the table. The astronomical searches for pyridine by different groups gave only the upper limits of its column density in the range of $7.3*10^{12}$ to $2.5*10^{15}$ cm$^{-2}$ without any successful detection. [16,17,18] As much as we know, there is no report on the astronomical search for other members of this isomeric group. Figure 3 is the optimized structure of pyridine while Figure 4 shows the optimized structures of the hydrogen bonded complexes of these isomers with water molecule on the surface of the interstellar dust grains. After azafulvene, the next most affected molecule by the effect of interstellar hydrogen bonding is our molecule of interest; pyridine suggesting that a portion of it is probably attached to the surface of the interstellar dust grains thereby reducing the chances of its successful astronomical detection.

Table 3: $C_5H_5N$ isomers, Binding Energy with water, $\Delta H^o_f$ and Dipole moment

| Molecule | [a]$\Delta H^o_f$ (kcal/mol) | B. E with $H_2O$ (kcal/mol) | Dipole moment (Debye) |
|---|---|---|---|
| Pyridine | 28.737 | -4.752 | 2.434 |
| 1-cyano-1,3-butadiene | 55.518 | -4.170 | 5.295 |
| Azafulvene | 56.062 | -5.564 | 2.174 |
| 2-methylene-3-butanenitrile | 58.171 | -3.811 | 4.212 |

[a] [ref. 30]

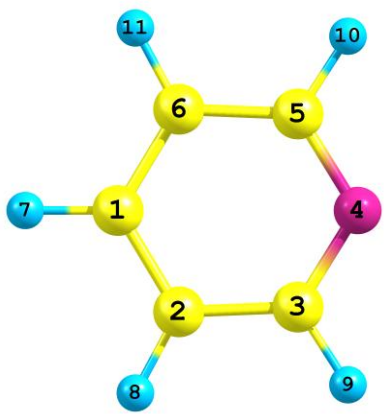

Figure 3: Optimized structure of Pyridine

The low dipole moment of pyridine in comparison to its isomers is also an issue of concern. Though 2-methylene-3-butanenitrile and 1-cyano-1,3-butadiene are thermodynamically less stable than pyridine but the less effect of interstellar hydrogen bonding on them and their high dipole moment are good factors that should warrant their astronomical searches. Their successful observations will not only add to the list of known interstellar and circumstellar molecules but will obviously increase the hope of detecting pyridine.

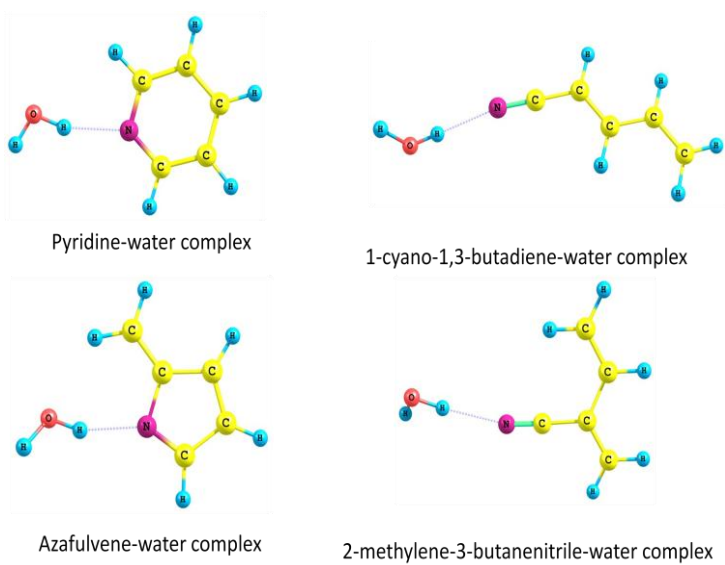

**Figure 4:** Optimized structures of the hydrogen bonded $C_5H_5N$ isomer complexes with water

As shown in Figure 4, in all the cases the water molecule acts as hydrogen bond donor with an elongation of one of its hydrogen bonds from the original 0.959Å in the monomer to 0.972 Å, 0.964 Å, 0.973 Å and 0.964 Å respectively in the pyridine, 1-cyano-1,3-butadiene, azafulvene and 2-methylene-3-butanenitrile-water complexes confirming the formation of the bond. Table 4 shows the ZPE of pyridine and its D-analogues, the Boltzmann factor (E/kT) at 10 and 100k; and the temperature where the Boltzmann factor is unity (*ΔT).

Table 4: ZPE and Boltzmann factor (E/kT) for Pyridine and its D-analogues

| Molecule | ZPE (kcal/mol) | *ΔT (K) | E/kT (at T=10k) | E/kT (at T=100k) |
|---|---|---|---|---|
| Pyridine | 55.655 | - | - | - |
| $C_5H_4D(7)N$ | 53.606 | 1031.339 | 103.134 | 10.313 |
| $C_5H_4D(8)N$ | 53.606 | 1031.339 | 103.134 | 10.313 |
| $C_5H_4D(9)N$ | 53.577 | 1045.549 | 104.555 | 10.455 |
| $C_5H_3D_2(8,9)N$ | 51.532 | 2074.993 | 207.499 | 20.7499 |
| $C_5H_2D_3(7,8,9)N$ | 49.481 | 3106.963 | 310.696 | 31.070 |
| $C_5HD_4(7,8,9,10)N$ | 47.402 | 4153.143 | 415.314 | 41.531 |
| $C_5D_5N$ | 45.354 | 5183.534 | 518.353 | 51.835 |

*The difference in the ZPE of a molecule and its D-analogue expressed in terms of temperature.

The number(s) in bracket is (are) as defined before. Unlike in the previous case where a pronounced difference in the ZPE was observed for singly substituted deuterium at different positions, such trend is not observed in the case of D-pyridine isotopomers. Deuterium fractionation is a temperature dependent process.[32] That a Boltzmann factor of unity is far from being achieved under conditions of the dense molecular clouds where these molecules are expected to be formed is a good omen. Since major deuterium fractionation occurs under the conditions where these molecules are formed, it thus implies that a high D/H ratio above the cosmic D/H ratio will be the result of such fractionation. Such high D/H ratio should of course prompt the astronomical search for D-pyridine.

## 3.3 Pyrimidine and Its Isomers

Table 5 shows the binding energy (B. E) of the $C_4H_4N_2$ isomers with water and their dipole moment. The zero-point corrected standard enthalpy of formation of these isomers form previous work is also included in the table.

Table 5: $C_4H_4N_2$ Isomers, Binding Energy with water, $\Delta H^o_f$ and Dipole moment

| Molecule | B. E with $H_2O$ (kcal/mol) | [a]$\Delta H^o_f$ (kcal/mol) | Dipole moment (Debye) |
|---|---|---|---|
| Pyrimidine | -4.490 | 37.114 | 2.523 |
| Pyrazine | -3.889 | 41.006 | 0.000 |
| 1,1-dicyanoethane | -3.285 | 54.713 | 4.657 |
| Pyridazine | -4.777 | 59.052 | 4.602 |

[a] [ref. 30]

Pyrimidine is the most stable isomer of the $C_4H_4N_2$ isomeric group. This has led to its astronomical searches with the column density of $7*10^{14}$, $2.4*10^{14}$, and $3.4*10^{14}$ cm$^{-2}$ determined for its upper limits in SgrB2(N), Orion KL and W51 respectively[19] without any successful detection from any of the astronomical sources. With respect to interstellar hydrogen bonding, pyridazine and pyrimidine are the most bonded isomers of the $C_4H_4N_2$ isomeric group to the surface of the interstellar dust grains. Thus suggests a possible reduction in the overall abundance of pyrimidine since a portion of it is strongly attached to the surface of the interstellar dust grains. Among these isomers, 1,1-dicyanoethane is the least affected by interstellar hydrogen bonding. This factor coupled with its high dipole moment and the high number of known cyanide molecules from interstellar space makes it a suitable candidate for astronomical search with a high expectation for possible observation. Figure 5 pictures the optimized structures of the hydrogen bonded complexes of these isomers with water. In the pyridazine-water complex where the two O-H bonds of the water monomer are taking part in the complex formation, there is an elongation of both bond lengths from the original 0.959Å to 0.962Å for each of the O-H bonds of the water monomer while in the cases of pyrimidine, pyrazine and 1,1-dicyanoethane-complexes, one of the O-H bonds elongates to 0.969 Å , 0.969 Å  and 0.963 Å  in that order.

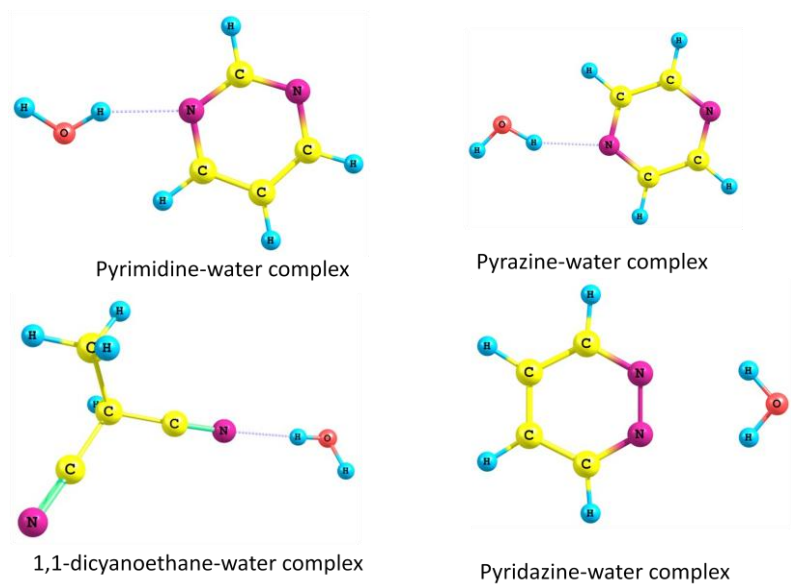

**Figure 5:** Optimized structures of the hydrogen bonded $C_4H_4N_2$ isomer complexes with water

Figure 6 is the structure of pyrimidine showing the numbering of the atoms as guide for the deuterium substitution at different positions as presented in Table 6. The H-atoms at positions 8 and 10 are identical. For the non-identical H-atoms, deuterium substitution at these positions leads to different values of ZPE which results in the different values of the Boltzmann factor for the singly deuterated pyrimidine. This implies different D/H ratio for these species in the interstellar space.

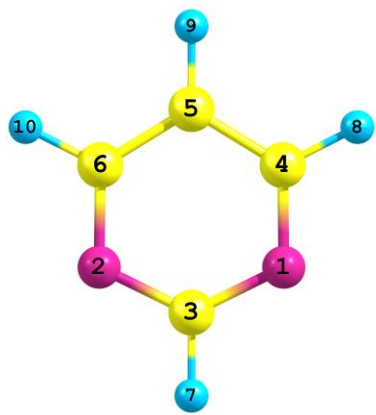

Figure 6: Optimized structure of Pyrimidine

From Table 6, major deuterium fractionation is expected to be observed for these D-analogues of pyrimidine since the prevailing conditions in the molecular clouds make it impossible to achieve a Boltzmann factor of unity. The high Boltzmann factor at the low temperature of the molecular clouds will thus favor high abundance of these deuterated molecules, thus their successful observation can be envisaged.

**Table 6:** ZPE and Boltzmann factor (E/kT) for Pyrimidine and its D-analogues

| Molecule | ZPE (kcal/mol) | *ΔT (K) | E/kT (at T=10k) | E/kT (at T=100k) |
|---|---|---|---|---|
| Pyrimidine | 48.279 | - | - | - |
| $C_4H_3D(7)N_2$ | 46.175 | 1059.127 | 105.913 | 10.591 |
| $C_4H_3D(8)N_2$ | 46.207 | 1043.022 | 104.302 | 10.430 |
| $C_4H_3D(9)N_2$ | 46.240 | 1025.970 | 102.597 | 10.260 |
| $C_4H_2D2(7,8)N_2$ | 44.101 | 2102.465 | 210.246 | 21.025 |
| $C_4HD_3(7,8,9)N_2$ | 42.062 | 3128.752 | 312.875 | 31.287 |
| $C_4D_4N_2$ | 39.988 | 4172.406 | 417.241 | 41.724 |

*The difference in the ZPE of a molecule and its D-analogue expressed in terms of temperature.

## 3.4 Pyrrole and Its Isomers

Unlike in the previous cases where only the most stable isomer is the there only molecule that has been astronomically searched for, in the $C_4H_5N$ isomeric group, both the most stable isomer (pyrrole) and a less stable isomer; ally isocyanide have been astronomically searched from different molecular clouds with only upper limits determined for both without any successful detection. The dipole moment, binding energy (B. E) of the complex formed between these isomers and water are captured in Table 7 including the zero-point corrected standard enthalpy of formation of these isomers from previous study. The astronomical search of more than one isomer from this group provides an excellent opportunity to practically test the effect of interstellar hydrogen bonding discussed here. Pyrrole (24.094kcal/mol) is extremely more stable as compared to ally isocyanide (61.986 kcal/mol). However, with respect to the effect of interstellar hydrogen bonding, pyrrole is the most bonded isomer of the $C_4H_5N$ isomeric group to the surface of the interstellar dust grains while ally isocyanide is the second least affected isomer as regards this effect.

Table 7: **$C_4H_5N$**Isomers, Binding Energy with water,$\Delta H^o_f$ and Dipole moment

| Molecule | B. E with $H_2O$ (kcal/mol) | [a]$\Delta H^o_f$ (kcal/mol) | Dipole moment (Debye) |
|---|---|---|---|
| Pyrrole | -4.654 | 24.094 | 1.933 |
| 2-butenenitrile | -4.245 | 32.787 | 5.244 |
| 2-cyanopropene | -3.073 | 34.031 | 4.436 |
| Ally cyanide | -4.0423 | 39.747 | 4.318 |
| Ally isocyanide | -3.214 | 61.986 | 3.597 |

[a] [ref. 30]

The column density determined for pyrrole ranges from 3 to$10*10^{13}$ cm$^{-2}$ while for ally isocyanide, 1.1 to $3.3*10^{14}$ cm$^{-2}$ is reported as its range of column density.[18,2] The high column density reported for ally isocyanide clearly shows that a large portion of pyrrole is strongly attached to the surface of the interstellar dust grains thus reducing its overall abundance. Considering 2-cyanopropene which is less bonded to the surface of the interstellar dust grains than every other isomer; its thermodynamic stability is almost doble that of ally isocyanide while its dipole moment is higher than those of pyrrole and ally isocyanide. These properties of 2-cyanopropene in addition to the reported column density of pyrrole and ally isocyanide put it in a right place for a successful astronomical observation.

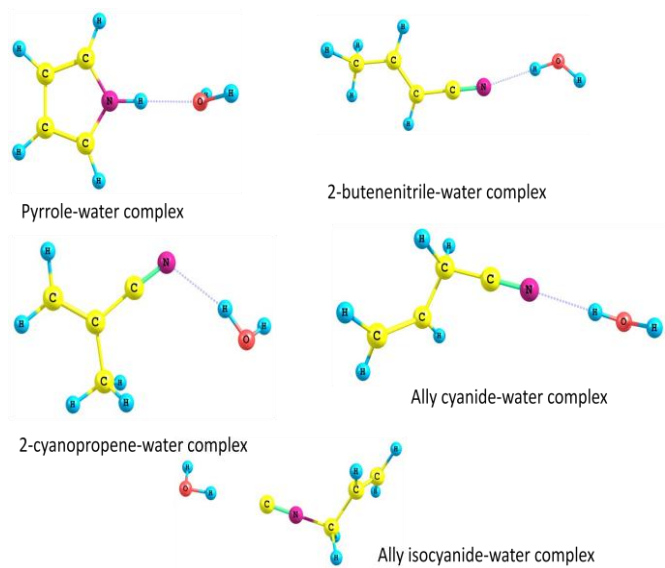

figure 7: Optimized structures of the hydrogen bonded $C_4H_5N$ isomer complexes with water

Figure 7 depicts the optimized structures of the hydrogen bonded $C_4H_5N$ isomer complexes with water with the water monomer acting as hydrogen bond donor in all the complexes here with an elongation of the O-H bond.

The non-identical H-atoms in pyrrole (as pictures in Figure 8) show similar effect on deuteration. Deuterium substitution at positions 7 and 10 results in exactly the same ZPE as it is also noted for substitution at positions 6 and 9 (Table 8).

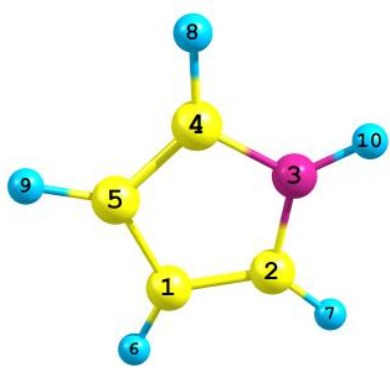

Figure 8: Optimized structure of Pyrrole

The large Boltzmann factor makes it clear that deuterated pyrrole occurs under the same conditions as its main isotopologue (low temperature interstellar cloud conditions). Under these conditions, pronounced deuterium fractionation from its reservoir will lead to an enhanced D/H ratio far above the cosmic D/H ratio which increases the probability of its astronomical observation.

**Table 8:** ZPE and Boltzmann factor (E/kT) for Pyrrole and its D-analogues

| Molecule | ZPE (kcal/mol) | *ΔT (K) | E/kT (at T=10k) | E/kT (at T=100k) |
|---|---|---|---|---|
| Pyrrole | 51.736 | - | - | - |
| $C_4H_4D(10)N$ | 49.580 | 1085.021 | 108.502 | 10.850 |
| $C_4H_4D(7)N$ | 49.580 | 1085.021 | 108.502 | 10.850 |
| $C_4H_4D(8)N$ | 49.724 | 1012.392 | 101.239 | 10.124 |
| $C_4H_4D(6)N$ | 49.724 | 1012.708 | 101.271 | 10.127 |
| $C_4H_4D(9)N$ | 49.724 | 1012.708 | 101.271 | 10.127 |
| $C_4H_3D_2(6,7)N$ | 47.719 | 2021.626 | 202.162 | 20.216 |
| $C_4H_2D_3(6,7,8)N$ | 45.713 | 3031.175 | 303.117 | 30.312 |
| $C_4HD_4(6,7,8,9)N$ | 43.706 | 4041.041 | 404.104 | 40.410 |
| $C_4D_5N$ | 41.554 | 5124.167 | 512.417 | 51.242 |

*The difference in the ZPE of a molecule and its D-analogue expressed in terms of temperature.

### 3.5 Quinoline, Isoquinoline and their Iosmers

Table 9 shows the six most stable isomers of the $C_9H_7N$ isomeric group. Unsuccessful astronomical searched have been reported for the two most stable isomers of the group; quinoline and isoquinoline.[18] The strong attachments of heterocycles to the surface of the interstellar dust grains as compared to their respective isomers observed in the previous cases is not left out here. As shown in Table 9, quinoline is the most strongly bonded isomer of the group to the surface of the interstellar dust grains while its closest counterpart; isoquinoline ranks third. This effect affects the interstellar abundance of these heterocycles thus influencing their successful detection. Figure 9 displays the optimized structures of the hydrogen bonded $C_9H_7N$ isomer complexes with water. The elongation of the O-H bond from its original length signifies bond formation.

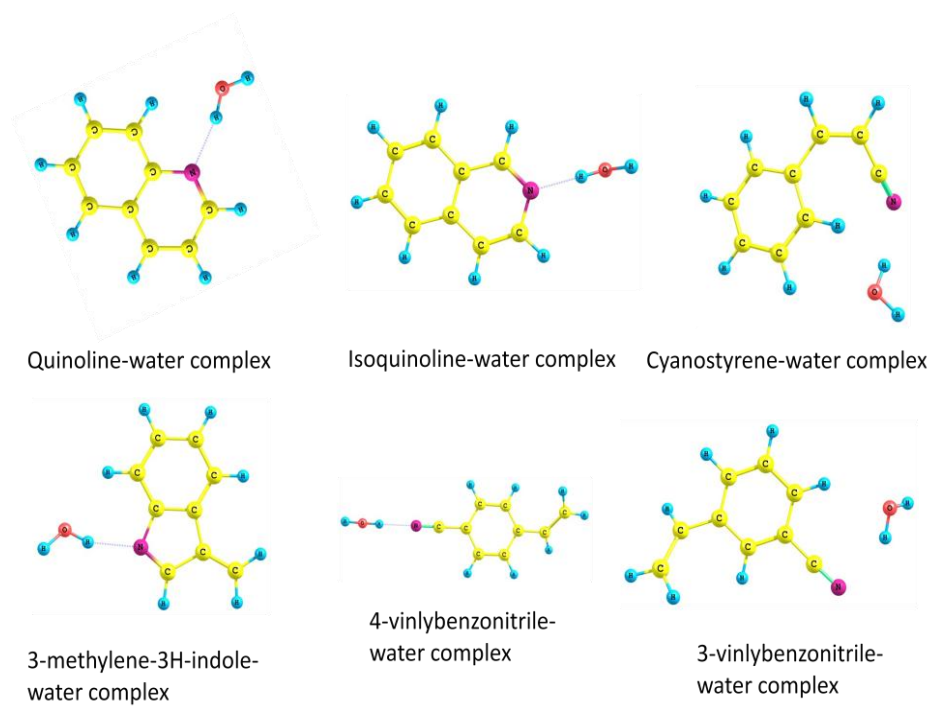

Figure 9: Optimized structures of the hydrogen bonded $C_9H_7N$ isomer complexes with water

Table 9: $C_9H_7N$ Isomers, Binding Energy with water, $\Delta H^o_f$ and Dipole moment

| Molecule | B. E with $H_2O$ (kcal/mol) | [a]$\Delta H^o_f$ (kcal/mol) | Dipole moment (Debye) |
|---|---|---|---|
| Quinoline | -4.239 | 44.781 | 2.305 |
| Isoquinoline | -3.937 | 45.956 | 2.730 |
| Cyanostyrene | -3.309 | 66.518 | 4.691 |
| 3-methylene-3H-indole | -4.025 | 66.643 | 2.948 |
| 4-vinylbenzonitrile | -3.640 | 66.918 | 5.402 |
| 3-vinylbenzonitrile | -3.390 | 67.432 | 5.125 |

[a] [ref. 30]

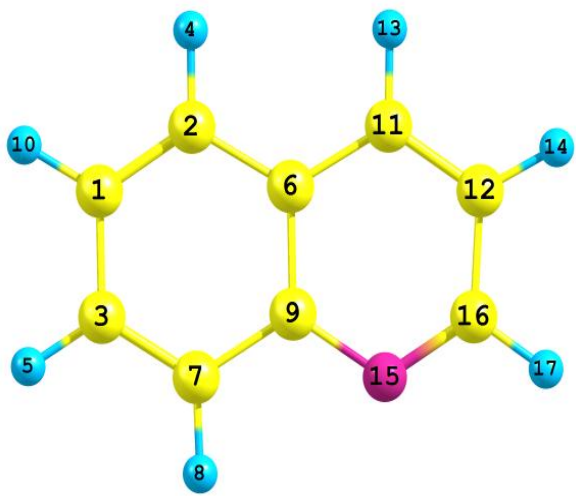

Figure 10: Optimized structure of Quinoline

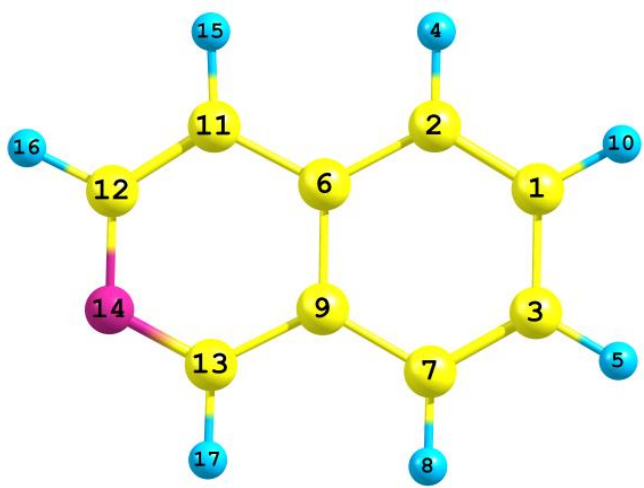

Figure 11: Optimized structure of Isoquinoline

Quinoline and isoquinoline belong to the nitrogen containing PAHs. The non-detection of fully hydrogenated PAHs and related compounds from the interstellar ice grains is partly linked to the loose of spectroscopic signature of these molecules due their transformations to more complex organics.[24] This places an important demand on the astronomical searches for the deuterated analogue of these molecules. The numbering in Figures 10 and 11 is as defined before. The Boltzmann factor shown in Tables 10 and 11 support major deuterium fractionation for the formation of the D-analogues of quinoline and isoquinoline. This implies the possible present of these deuterated molecules in detectable amount.

Table: 10: ZPE and Boltzmann factor (E/kT) for Quinoline and its D-analogues

| Molecule | ZPE(kcal/mol) | *ΔT(K) | E/kT (at T=10k) | E/kT(at T=100k) |
|---|---|---|---|---|
| Quinoline | 84.948 | - | - | - |
| $C_9H_6D(17)N$ | 82.871 | 1045.233 | 104.523 | 10.452 |
| $C_9H_6D(14)N$ | 80.120 | 2429.614 | 242.961 | 24.296 |
| $C_9H_6D(13)N$ | 82.904 | 1028.497 | 102.850 | 10.285 |
| $C_9H_6D(8)N$ | 82.901 | 1030.075 | 103.007 | 10.301 |
| $C_9H_5D_2(14,17)N$ | 80.822 | 2076.256 | 207.626 | 20.763 |
| $C_9H_5D_2(5,10)N$ | 80.842 | 2066.151 | 206.615 | 20.661 |
| $C_9H_4D_3(13,14,17)N$ | 78.772 | 3107.594 | 310.759 | 31.076 |
| $C_9D_7N$ | 70.563 | 7238.633 | 723.863 | 72.386 |

*The difference in the ZPE of a molecule and its D-analogue expressed in terms of temperature.

Table 11: ZPE and Boltzmann factor (E/kT) for Isoquinoline and its D-analogues

| Molecule | ZPE(kcal/mol) | *ΔT(K) | E/kT (at T=10k) | E/kT(at T=100k) |
|---|---|---|---|---|
| Isoquinoline | 84.978 | - | - | - |
| $C_9H_6D(17)N$ | 82.911 | 1040.18 | 104.018 | 10.402 |
| $C_9H_6D(16)N$ | 82.900 | 1045.864 | 104.586 | 10.459 |
| $C_9H_6D(15)N$ | 82.939 | 1025.97 | 102.597 | 10.260 |
| $C_9H_6D(4)N$ | 80.518 | 2244.251 | 224.425 | 22.442 |
| $C_9H_5D_2(16,17)N$ | 80.827 | 2088.887 | 208.889 | 20.889 |
| $C_9H_4D_3(15,16,17)N$ | 78.783 | 3117.383 | 311.738 | 31.174 |
| $C_9H_4D_3(4,8,10)N$ | 78.828 | 3094.647 | 309.465 | 30.946 |
| $C_9D_7N$ | 70.572 | 7249.685 | 724.968 | 72.497 |

*The difference in the ZPE of a molecule and its D-analogue expressed in terms of temperature.

### 3.6 Furan and Its Isomers

The $C_4H_4O$ isomeric group is the only O-containing group considered in this study. Furan, the most stable isomer of this series is a cyclic molecule with 9 atoms. Though there is no cyclic molecule with 9 atoms among the known interstellar and circumstellar molecules, about 10 linear molecules with 9 atoms have so far been detected from different astronomical sources.[3] The detection of furan will be an interesting development, firstly as a heterocyclic molecule and secondly as an addition to the very few known cyclic interstellar and circumstellar molecules. Unlike in the previous cases where the heterocyclic molecule always falls among the top most bonded isomer to the surface of the interstellar dust grains, the trend is reversed here. Furan is the second least affected isomer with respect to interstellar hydrogen bonding as shown by the data in Table 12. This suggests that a greater portion of furan is not attached to the surface of the interstellar dust grains, thus its overall abundance is not largely affected. The report on the astronomical searches for furan came in the last 3 to 4 decades[15,20] with its upper limits of column density determined to be $7*10^{13}$ cm$^{-2}$ and $2*10^{16}$ cm$^{-2}$ in Orion A and Sgb B2 respectively. Considering the advances in astronomical instruments and the complex molecules that have been astronomically detected within this period, it suffices to say that a new dedicated search for interstellar furan will probably be successful.

Table 12: $C_4H_4O$ Isomers, Binding Energy with water, $\Delta H^o_f$ and Dipole moment

| Molecule | B.E with $H_2O$ (kcal/mol) | [a]$\Delta H^o_f$ (kcal/mol) | Dipole moment (Debye) |
|---|---|---|---|
| Furan | -2.551 | -9.261 | 0.851 |
| Vinylketene | -1.954 | 3.563 | 1.774 |
| 2-cyclobutene-1-one | -4.826 | 9.180 | 4.166 |
| 2,3-butadienal | -4.525 | 13.116 | 4.367 |

[a] [ref. 30]

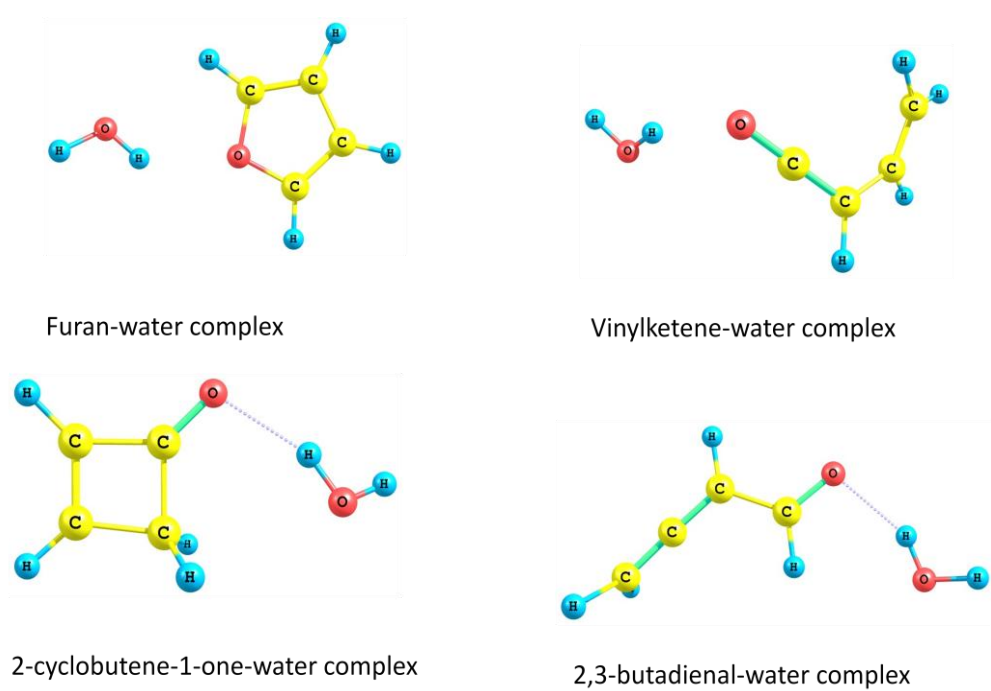

Furan-water complex

Vinylketene-water complex

2-cyclobutene-1-one-water complex

2,3-butadienal-water complex

Figure 12: Optimized structures of the hydrogen bonded $C_4H_4O$ isomer complexes with water

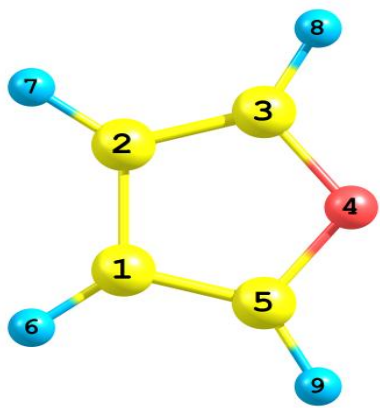

Figure 13: Optimized structure of Furan

Figure 12 nicely pictures the optimized structures of the hydrogen bonded $C_4H_4O$ isomer complexes with water monomer on the surface of the interstellar dust grains. In all the cases, the water molecule acts as hydrogen bond donor with an elongation for the O-H bond from its original 0.959Å to 0.962Å, 0.961Å, 0.966Å and 0.966Å in furan, vinylketene, 2-cyclobutene-1-one and 2,3-butadienal respectively indicating the bond formation. The numbering in Figure 13 stands as earlier defined. As presented in Table 13, the parameters for the deuterated analogue of furan follow the same trend as the cases discussed above with major deuterium fractionation being the dominant process for the formation of D-furan.

Table 13: ZPE and Boltzmann factor (E/kT) for Furan and its D-analogues

| Molecule | ZPE(kcal/mol) | *ΔT(K) | E/kT (at T=10k) | E/kT(at T=100k) |
|---|---|---|---|---|
| Furan | 43.852 | - | - | - |
| $C_4H_3D(8)O$ | 41.825 | 1020.286 | 102.029 | 10.203 |
| $C_4H_3D(9)O$ | 41.825 | 1020.286 | 102.029 | 10.203 |
| $C_4H_3D(6)O$ | 41.850 | 1007.655 | 100.765 | 10.076 |
| $C_4H_3D(7)O$ | 41.850 | 1007.655 | 100.765 | 10.076 |
| $C_4H_2D2(8,9)O$ | 39.796 | 2041.204 | 204.120 | 20.412 |
| $C_4H_2D2(6,7)O$ | 39.846 | 2015.942 | 201.594 | 20.159 |
| $C_4HD_3(6,7,8)O$ | 37.816 | 3037.491 | 303.749 | 30.375 |
| $C_4D_4O$ | 35.786 | 4059.356 | 405.936 | 40.594 |

*The difference in the ZPE of a molecule and its D-analogue expressed in terms of temperature.

## 4. Conclusions

Effect of interstellar hydrogen bonding on the astronomical detection of heterocycles and the possible detectability of the deuterated analogues of these heterocycles have been examined as strategies to optimize the astronomical searches for these heterocycles. On the effect of interstellar hydrogen bonding, imidazole, pyridine, pyrimidine, pyrrole, quinoline and isoquinoline are found to be strongly bonded to the surface of the interstellar dust grains as compared to their other stable isomers. This causes a greater portion of these molecules to be attached to the surface of the interstellar dust grains thereby reducing their overall abundance thus influencing their successful astronomical observation. On the basis of this, the other isomers of these heterocycles could be observed despite their low thermodynamic stability. Successful detection of furan remains highly feasible as it is not really affected by this effect. With respect to the detectability of the deuterated analogues of these heterocycles, the Boltzmann factor determined for all the D-analogues of these molecules at the temperature range of the dense molecular cloud are far above unity which implies major deuterium fractionation as the formation process for these D-analogues. This will result in very high D/H ratio far above the cosmic D/H ratio implying a high abundance of the D-analogue which is desirable for their successful detection.

**Author Contributions**

The manuscript was written through the contributions of all authors. E.E Etim designed the study.

ORCID [ID]
E. E. Etim
C. Andrew      0000-0003-2481-2225

**Acknowledgements:** EEE acknowledges a research fellowship from the Indian Institute of Science, Bangalore.